\documentstyle[aps,twocolumn,prb,epsf]{revtex}
\begin{document}
\title{Negative magnetoresistance in a mixed-valence
$La_{0.6}Y_{0.1}Ca_{0.3}MnO_3$:\\
Evidence for charge localization governed by the Curie-Weiss law}
\author{S. Sergeenkov$^{1,2}$, H. Bougrine$^{1,3}$, M. Ausloos$^{1}$, and
A. Gilabert$^{4}$}
\address{$^{1}$SUPRAS, Institute of Physics, B5, University of Li$\grave e$ge,
B-4000 Li$\grave e$ge, Belgium\\
$^{2}$Bogoliubov Laboratory of Theoretical Physics,
Joint Institute for Nuclear Research,\\ 141980 Dubna, Moscow Region, Russia\\
$^{3}$SUPRAS, Montefiore Electricity Institute, B28, University of
Li$\grave e$ge, B-4000 Li$\grave e$ge, Belgium\\
$^{4}$Laboratoire de Physique de la Mati$\grave e$re Condens\'ee, Universit\'e
de Nice-Sophia Antipolis,\\ Parc Valrose F-09016 Nice, Cedex 02, France\\}
\date{\today}
\draft
\maketitle
\begin{abstract}
Colossal negative magnetoresistance $\Delta \rho (T,B)=\rho (T,B)-\rho (T,0)$
observed in
$La_{0.6}Y_{0.1}Ca_{0.3}MnO_3$ at $B=1T$ shows a nearly perfect
symmetry around $T_{0}=160K$
suggesting a universal field-induced transport mechanism in this
material. Attributing this symmetry to strong magnetic fluctuations
(triggered by the $Y$ substitution and further enhanced by magnetic field,
both above and below the field-dependent Curie temperature $T_{C}(B)\equiv T_0$),
the data are interpreted in terms of the nonthermal spin hopping and
magnetization $M$ dependent
charge carrier localization scenario leading to
$\Delta \rho =-\rho _s\left( 1-e^{-\gamma M^{2}}\right )$
with $M(T,B)=CB/|T-T_C|^{\nu}$.
The separate fits through all the data points above and below $T_C$
yield $C^{+}\simeq C^{-}$ and $\nu ^{+}\simeq \nu ^{-}\simeq 1$. The
obtained results corroborate the importance of
fluctuation effects in this material recently found to dominate its
magneto-thermopower behavior far beyond $T_C$.
\pacs{PACS numbers: 72.15.Gd, 71.30.+h, 75.70.Pa}
\end{abstract}

\narrowtext

Since recently, interest in the mixed-valence manganite perovskites
$R_{1-x}Ca_{x}MnO_3$
(where $R=La,Y,Nd,Pr$) has been renewed due to the large negative
magnetoresistive (MR) effects observed near the ferromagnetic (FM) ordering
of $Mn$ spins.~\cite{1,2,3,4,5,6,7,8,9,10,11}
In the doping range $0.2<x<0.5$, these compounds are known to undergo a double
phase transition from paramagnetic-insulating to
ferromagnetic-metallic state near the Curie temperature $T_C$.
Above $x=0.5$, the specific heat and susceptibility measurements
reveal~\cite{6,7} an extra antiferromagnetic (AFM) canted-like transition
at $T_{AFM}$ lying below $T_C$. At the same time, substitution
on the $La$ site was found to modify the phase diagram through cation size
effects leading toward either charge-ordered (CO) or AFM instability.~\cite{6}
In particular, $Y$ substitution is responsible for two major modifications
of the parent manganite: (i) it lowers the FM Curie temperature $T_C$, and
(ii) weakens the system's robustness against strong AFM fluctuations (which
are developed locally within the ordered FM matrix) by shifting $T_C$ closer
to $T_{AFM}$. The latter is considered~\cite{8,9} as the most probable
reason for strong magnetic localization of spin polarized carriers (forming
the so-called spin polarons) which in
turn results in hopping dominated charge carrier transport mechanism below
$T_C$. While above $T_C$, the resistivity presumably follows a thermally
activated Mott-like variable-range hopping
law $\rho \propto \exp(T_0/T)^{z}$ with $1/4\leq z \leq 1$.
On the other hand, there are some indications~\cite{10} that the observable MR
$\Delta \rho (T,B)$ scales with magnetization $M$ in the ferromagnetic state
and follows $M^{2}$ dependence in the paramagnetic region implying thus
some kind of universality in the magnetotransport behavior of (low
conductive) manganites below and above $T_C$.

In this paper we present some typical results
for magnetoresistivity (MR) measurements on a manganite sample
$La_{0.6}Y_{0.1}Ca_{0.3}MnO_3$ at $B=1T$ field for a wide temperature
interval (ranging from $20K$ to $300K$) and compare them with the available
theoretical explanations. As we shall see, the data are best described
in terms of the nonthermal (rather than Mott-like thermally assisted) spin
hopping scenario with magnetization- dependent
charge carrier localization length $L(M)$ both above and below $T_C$.
The interpretation is essentially based on the assumption of rather
strong magnetic fluctuations in this material far beyond the Curie point
$T_C$ (see also the previously reported~\cite{12} discussion of the
magneto-thermopower).

The polycrystalline $La_{0.6}Y_{0.1}Ca_{0.3}MnO_3$ samples used in our
measurements were prepared
from stoichiometric amounts of $La_{2}O_3$, $Y_{2}O_3$, $CaCO_3$, and $MnO_2$
powders.
The mixture was heated in air at $800C$ for 12 hours to achieve
the decarbonation and then pressed at room temperature to obtain
parallelipedic pellets. An annealing and sintering
from $1350C$ to $800C$ was made slowly (during 2 days) to preserve the
right phase stoichiometry. A small bar ($10mm\times 4mm^2$)
was cut from one pellet. The electrical resistivity $\rho (T,B)$ was
measured using the conventional four-probe method. To avoid Joule and
Peltier effects, a dc current $I=1mA$ was injected (as a one second pulse)
successively on both sides of the sample. The voltage drop $V$ across the
sample was measured with high accuracy by a $KT256$ nanovoltmeter. The
magnetic field $B$ of $1T$ was applied normally to the current.
Fig.1 presents the temperature dependence of the resistivity
$\rho (T,B)$ for a $La_{0.6}Y_{0.1}Ca_{0.3}MnO_3$
sample at zero and $B=1T$ field. The corresponding MR
$\Delta \rho (T,B)=\rho (T,B)-\rho (T,0)$ is shown in Fig.2 as a function
of reduced temperature
$(T-T_0)/T_0$ with $T_0=160K$ being the temperature where the negative MR
exhibits a minimum. Notice the nearly perfect symmetry of the MR with
respect to
left ($T<T_0$) and right ($T>T_0$) wings thus suggesting a "universal"
magnetotransport mechanism above and below $T_0$.
\begin{figure}[htb]
\epsfxsize=8cm
\centerline{\epsffile{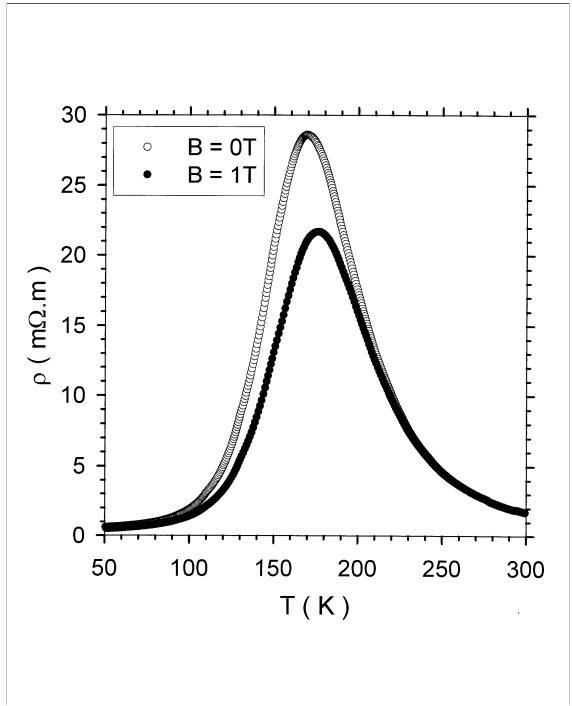} }
\caption{Temperature behavior of the observed resistivity
$\rho (T,B)$ in $La_{0.6}Y_{0.1}Ca_{0.3}MnO_3$ at zero and
$B=1T$ field.}
\end{figure}
Before discussing a
probable scenario for the observed MR temperature behavior, let us briefly
review the current theoretical models.
Several~\cite{8,9,10,11} unification approaches have been suggested.
In essence,
all of them are based on a magnetic localization
concept which relates the observable MR at any temperature and/or applied
magnetic field to the local magnetization.
In particular, one of the most
advanced models of this kind~\cite{9}
ascribes the metal-insulator (M-I) like transition
to a modification of the spin-dependent potential $J_H \vec s \cdot \vec S$
associated with the onset of magnetic order at $T_C$ (here $J_H\simeq 1eV$ is
the
on-site Hund's-rule exchange coupling of an $e_g$ electron with $s=1/2$ to
the localized $Mn$ $t_{2g}$ ion core with $S=3/2$). Specifically, the
hopping based conductivity reads
\begin{equation}
\sigma =\sigma _m\exp \left (-\frac{2R}{L}-\frac{W_{ij}}{k_BT}\right ),
\end{equation}
where
\begin{equation}
\sigma _m=e^{2}R^{2}\nu _{ph}N(E_m).
\end{equation}
Here $R$ is the hopping distance (typically,~\cite{10} of the order of $1.5$
unit cells),
$L$ the charge carrier localization length (typically,~\cite{10} $L\simeq 2R$),
$\nu _{ph}$ the phonon frequency, $N(E_m)$ the density of available states
at the magnetic energy $E_m\simeq J_H$, and
$W_{ij}$ the effective barrier between the hopping sites $i$ and $j$.
There are two possibilities to introduce an explicit magnetization dependence
into the above model: either by modifying the hopping barrier $W_{ij}\to
W_{ij}-\alpha \vec M_i \cdot \vec M_j$ or by assuming a magnetization-dependent
localization length $L(M)$. The first scenario (suggested
by Viret {\it et al.}~\cite{9}) results in a thermally activated behavior of
MR over the whole temperature range. Indeed, since
a sphere of radius $R$ contains $(4/3)\pi R^{3}/v$ sites where
$v=5.7\times 10^{-29}m^{3}$ is the lattice volume per manganise ion, the
smallest value of $W_{ij}$ is therefore $[(4/3)\pi R^{3}N(E_m)]^{-1}$.
Minimizing the hopping rate, one finds that the resistivity should vary as
$\ln(\rho /\rho _0)=[T_0\{1-(M/M_s)^{2}\}/T]^{1/4}$. 
\begin{figure}[htb]
\epsfxsize=8cm
\centerline{\epsffile{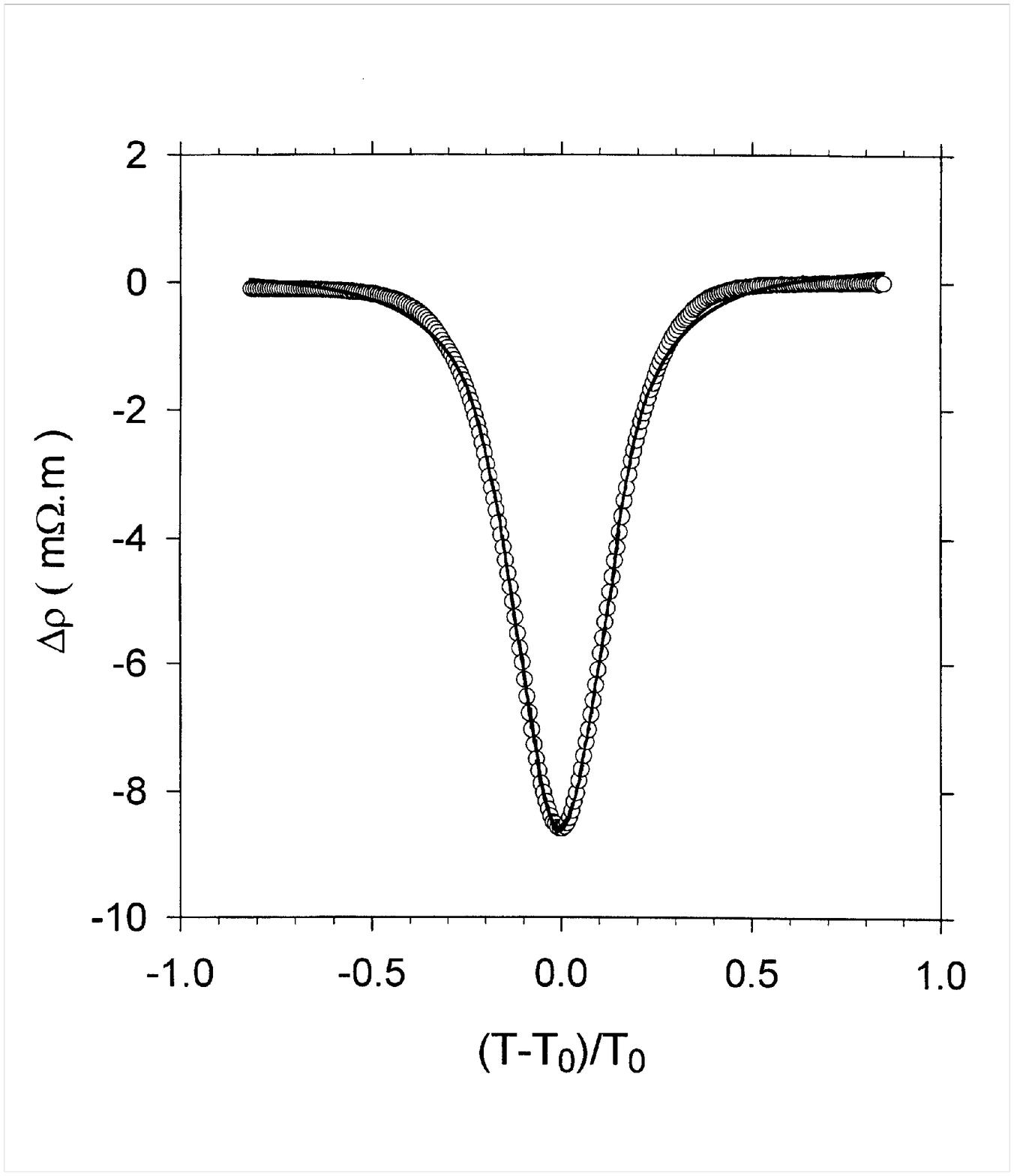} }
\caption{The dependence of the observed magnetoresistivity (MR)
$\Delta \rho (T,B)=\rho (T,B)-\rho (T,0)$ in
$La_{0.6}Y_{0.1}Ca_{0.3}MnO_3$ at $B=1T$ as a function of
$(T-T_0)/T_0$ with $T_0=160K$ being the temperature where the MR reaches
its minimum. The solid line through all the data points is the best fit
according to Eq.(3). }
\end{figure}
This scenario was used
by Wagner {\it et al.}~\cite{10} to interpret their GMR data on low-conductive
$Nd_{0.52}Sr_{0.48}MnO_3$ films.
Assuming the molecular-field result for the magnetization, they found that their
GMR data scale with the Brillouin function $\cal B$ in the ferromagnetic state
and follow a ${\cal B}^{2}$ dependence in the paramagnetic state.
Unfortunately, all attempts to fit our MR data with the above thermally activated
hopping formula have failed. In our case
the observed MR seems to follow a steeper temperature behavior exhibiting
a remarkable symmetry around $T_0$.
Instead, we were able to successfully fit our data for
the whole temperature interval with
\begin{equation}
\Delta \rho (T,B)=-A[1-e^{-\beta (T)}],
\end{equation}
where
\begin{equation}
\beta (T)=\beta _0\left[ \frac{T_0}{T-T_0}\right ]^{2\nu},
\end{equation}
and $A$, $\beta _0$ and $\nu$ are
temperature-independent parameters. The separate fits for our MR data above
and below
$T_0=160K$ produce $A=0.873\pm 0.001\Omega cm$, $\beta _0^{+}=0.015\pm 0.001$,
$\beta _0^{-}=0.016\pm 0.001$, $\nu ^{+}=0.98\pm 0.01$, and
$\nu ^{-}=1.01\pm 0.01$, in agreement with the observed symmetry.
This suggests us to interpret our findings in terms of a nonthermal
localization scenario,~\cite{11} which emphasizes the role
of a nonmagnetic disorder and assumes a magnetization dependence
of the localization length $L(M)$ (rather than hopping barrier $W_{ij}$)
which diverges at the M-I phase transition.
Within this scenario, the Curie point $T_C$
is defined through the Curie-Weiss susceptibility $\chi =C/(T_C-T)$ as
$\chi ^{-1}(T_C,B)=0$,
while the M-I transition temperature $T_{MI}$ is such that $M(T_{MI},B)=M_0$
(with $M_0$ being a fraction of the saturated magnetization $M_s$).
In terms of the spontaneous magnetization $M$,
it means that for $M<M_0$ the system is in a highly resistive (insulator-like)
phase, while for $M>M_0$ the system is in a low resistive (metallic-like)
state. Besides, it is worthwhile to mention that we used this model in
successfully describing the magneto-thermopower (TEP) data
on the same sample.~\cite{12}
Adopting this scenario (with $W_{ij}/k_BT\ll 2R/L$, see Eq.(1)), we can
write $\rho (T,B)=\rho _0(T)+\rho _m\exp[2R/L(M)]$
for the field-induced resistivity in our sample. Here, $\rho _0(T)$ is a
field-independent background resistivity, $\rho _m=1/\sigma _m$, and
the localization length $L(M)$ depends on the field and temperature
through the corresponding dependencies of the magnetization $M(T,B)$.
Assuming after Sheng {\it et al.}~\cite{11} that $L(M)=L_0/(1-M^{2}/M_{0}^{2})$,
we arrive at the following simple expression for the MR
\begin{equation}
\Delta \rho (T,B)=-\rho _s\left[ 1-e^{-\gamma M^{2}(T,B)}\right ],
\end{equation}
where $\gamma =2R/L_0M_{0}^{2}$ and $\rho _s=\rho (T,0)-\rho _0(T)$ is the
temperature-independent residual resistivity. In fact, as we shall see
$\rho _s=\Delta \rho (T_0,B)$. To account for the observed symmetry of
the MR around $T_0$, we identify $T_0$ with the Curie temperature $T_C(B)$
at the finite magnetic field $B$, and
assume that the field and temperature dependence of the magnetization is
governed
by the same Curie-Weiss like law $M(T,B)=\chi (T,B)B$ with $\chi (T,B)=
C/|T-T_C|^{\nu}$ both above and below $T_C$.
Given the above definitions, we obtain $|T_C-T_{MI}|^{\nu}=CB/M_0$ for the
difference between the two critical temperatures which implies that within
the Curie-Weiss scenario, $T_{MI}=T_C(0)$. Finally, by comparing Eq.(5) with
the above-used fitting formula (see Eq.(3)), we arrive at the following
relations between
the fitting and model parameters, viz., $A=\rho _s$, $\beta _0=
(2R/L_0)[1-T_C(0)/T_C(B)]^{2\nu}$, and $T_0=T_C(B)$. Taking into account
the zero-field Curie temperature for this material~\cite{5} $T_C(0)=144K$ and
the found values of $\beta _0\simeq 0.015$ and $\nu \simeq 1$ we obtain
$2R/L_0\simeq 1$ for the
hopping distance to localization length ratio. Furthermore, knowing
this ratio and using
the found value of the residual resistivity $\rho _s\simeq 0.873\Omega cm$
gives an estimate for the hopping distance $R$ provided the density of states
$N(E_m)$ and the phonon frequency $\nu _{ph}$ are known. Using~\cite{9}
$N(E_m)\simeq
9\times 10^{26}m^{-3}eV^{-1}$ and $\nu _{ph}\simeq 2\times 10^{13}s^{-1}$
(estimated from Raman shift for optical $Mn-O$ modes~\cite{10}) for these two
parameters,
we arrive at a reasonable value of $R\simeq 5.5\AA$ which in turn
results in $L_0\simeq 11\AA$ for a zero-temperature zero-field carrier charge
localization length, in good agreement with the other
reported~\cite{3,5,7,8,9,10,11} estimates
of this parameter (as well as the one deduced from our own magneto-TEP
measurements~\cite{12}).

In conclusion, we would like to comment on the plausibility of our
interpretation which is essentially based on the Curie-Weiss behavior of
magnetization. Clearly, the possibility to use the Curie-Weiss law (which is
usually limited to the critical region around $T_C$)
throughout the whole region (ranging from the paramagnetic
to the ferromagnetic state) suggests the presence of strong Gaussian
(rather than critical) fluctuations both above and below $T_C$.~\cite{13}
To account for a possible source of these fluctuations, we turn to the
magnetic structure of our sample. As we
mentioned in the introductory part, along with lowering the Curie point,
$Y$ substitution brings about another important effect. Namely, it drives
the magnetic structure closer to a canted AFM phase (which
occurs~\cite{7} at $T_{AFM}<T_C$) thus triggering the development of local AFM
fluctuations (further enhanced by magnetic field) within the parent FM matrix.
In turn, these
fluctuations cause a trapping
of spin polarized carriers in a locally FM environment leading to
hopping dominated transport of charge carriers between thus
formed spin polarons, for the whole temperature interval.
And finally, it is worth noting that rather strong magnetic
fluctuation effects have been found to be responsible for the recently
observed~\cite{12} temperature behavior of the magneto-TEP for the same
sample as well, amounting to
$67\%$ and $33\%$ above and below $T_C$, respectively.

We thank J. C. Grenet and R. Cauro (University of Nice-Sophia Antipolis)
for lending us the sample.
Part of this work has been financially supported by the Action de Recherche
Concert\'ees (ARC) 94-99/174. M.A. and A.G. thank CGRI for financial support
through the TOURNESOL program.
S.S. thanks FNRS (Brussels) for some financial support.

\end{document}